\documentclass[referee,sn-basic]{sn-jnl}% referee option is meant for double line spacing

\usepackage{graphicx}%
\usepackage{subfigure}%
\usepackage{multirow}%
\usepackage{amsmath,amssymb,amsfonts}%
\usepackage{amsthm}%
\usepackage{mathrsfs}%
\usepackage[title]{appendix}%
\usepackage{xcolor}%
\usepackage{textcomp}%
\usepackage{manyfoot}%
\usepackage{booktabs}%
\usepackage{algorithm}%
\usepackage{algorithmicx}%
\usepackage{algpseudocode}%
\usepackage{listings}%
\begin{document}

\title[Long-term behavior of the optical polarization from the blazar 1ES 1959+650]{Insights into the Long-term behavior of the optical polarization from the blazar 1ES 1959+650}

\author*[1,2]{\fnm{K.~K.} \sur{Singh}}\email{kksastro@barc.gov.in}

\author[2]{\fnm{A.} \sur{Singh}}

\author[1,2]{\fnm{A.} \sur{Tolamatti}}

\author[3]{\fnm{P.~J.} \sur{Meintjes}}

\author[1,2]{\fnm{K.~K.} \sur{Yadav}}

\affil*[1]{\orgdiv{Astrophysical Sciences Division}, \orgname{Bhabha Atomic Research Centre}, \city{Mumbai}, \postcode{400085}, \country{India}}

\affil[2]{\orgname{Homi Bhabha National Institute}, \city{Mumbai}, \postcode{400094}, \country{India}}

\affil[3]{\orgdiv{Department of Physics}, \orgname{University of the Free State}, \city{Bloemfontein}, \postcode{9300}, \country{South Africa}}

%%-------------------------------------------------------Abstract---------------------------------------------------------------------
\abstract{A high degree of linear polarization measured in the optical emission is an important observational feature of blazars. 
It provides strong evidence of the presence of relativistic particles and magnetic field ordering in the non-thermal 
emission regions of blazars owing to the synchrotron nature of low energy radiation. Thus, the polarization studies of blazars 
are emerging as a promising approach to probe the particle acceleration and the physical processes involved in their broadband 
emission. In this work, we investigate the behavior of the optical polarization of the blazar 1ES 1959+650 measured over a 
decade using the spectropolarimetry (SPOL) at the Steward Observatory. We use measurements of the degree of linear polarization 
and angle of polarization in the wavelength range 500 - 700 nm available during the period October 1, 2008 and June 30, 2018 
(MJD 54739 - 58299) from the SPOL observations. Near simultaneous photometry data in the R and V bands are also used to study 
the optical emission from the source. The maximum degree of linear polarization, measured as $\sim$ 8.5$\%$, is significantly larger 
than the long term average value of $\sim$ 4.6$\%$. Analysis of the light curves indicates that the optical emission from the 
blazar 1ES 1959+650 is highly variable and variability in the degree of linear polarization can be quantified by a fractional 
variability amplitude of $\sim$ 39$\%$ over the period of about ten years. Long term optical emission in the R and V bands is 
very weakly anti-correlated with the degree of linear polarization. Modelling of the polarization due to the synchrotron emission 
suggests that the observed degree of linear polarization can be broadly reproduced by a power law distribution of relativistic 
electrons gyrating in a spherical emission region permeated with chaotic and ordered magnetic fields. Variation in the measured 
degree of polarization may be attributed to the interplay between the two magnetic field components in the emission region. 
The effect of stellar emission from the host galaxy of the blazar 1ES 1959+650 on the degree of synchrotron polarization originating 
from the jet is also discussed.}


\maketitle
%%--------------------------------------------------Section-1:Introduction-------------------------------------------------------------
\section{Introduction}\label{sec:Intro}
The phenomenon of blazars in the extragalactic universe is generally described by the accretion of matter onto a supermassive black hole 
and occurrence of a pair of oppositely directed sub-parsec to megaparsec scale jets with one of them pointing along the line of sight of 
the observer \citep{Urry1995,Padovani2017,Saikia2022}. These jets are characterized as the relativistic plasma outflows either in the 
continuous manner or in the form of discrete blobs originating from the vicinity of the black hole located at the center of the host galaxy. 
The azimuthal component of an initially poloidal magnetic field, caused by the accretion process, plays an important role in the formation 
of the relativistic jets \citep{Blandford1977,Tchekhovskoy2016,Chen2023a,Chen2023b}. This suggests that a helical magnetic field, ordered 
at parsec-scale distances, should be present in the jet. As a result, the blazar jets are observed to be the most powerful and persistent 
emitters of the non-thermal radiation in the universe \citep{Blandford2019}. The radiation, originating from the blazar jets, is relativistically 
beamed towards the observer at Earth and spans over the entire electromagnetic spectrum ranging from low-frequency radio to very high energy 
(VHE, above 100 GeV) $\gamma$-rays \citep{Urry1995,Blandford1978,Aharonian2008}. The broadband emission of the blazars exhibits some extreme 
observational properties such as the sudden flaring episodes with rapid and high amplitude flux variability at the diverse timescales ranging 
from minutes to years, strong and variable polarization, sharp variations in the polarization angle on different timescales, and superluminal 
motion \citep{Moore1981,Romero2000,Zhang2008,Abdo2010,Singh2015,Singh2018,Singh2019a,Hovatta2019,Raiteri2019,Singh2020a,Raiteri2021,Yang2022}.
\par
The origin of strong linear polarization up to $\sim$ 30$\%$ (theoretically expected maximum value $\sim 70\%$), measured in the radio and 
optical wavebands from a number of blazars is attributed to the synchrotron emission of relativistic electrons in a partially ordered jet 
magnetic field \citep{Westfold1959,Mead1990,Zhang2014,Fraija2017,Agudo2018,Singh2019b,Blinov2021}. The acceleration of electrons or other 
massive charge particles to the relativistic energies in the blazar jets is not fully understood and still remains an open problem. However, 
it is assumed that the particle acceleration is related to the jet magnetic field and can be probed through the polarization studies. 
The shock propagation through a turbulent medium having random magnetic field, change in blazar jet speed in a helical magnetic field, and relative 
orientation of the jets are among the popular hypotheses generally invoked to explain the observed features in the polarization light curves 
of blazars \citep{Marscher2014,Lyutikov2017}. The broadband non-thermal emission is calculated by taking into account the escape, radiative cooling and 
injection of the relativistic electrons in emission region, which is generally assumed to be compact single or multizone blobs (with different geometry 
like spherical, cylindrical, conical etc.) close to the base of the jet and moving relativistically along the jet. Single zone models, based on the relativistic 
electrons, have been greatly successful in reproducing the broadband spectral energy distributions (SEDs) of the blazars, however, existence of more than one 
emission region in the jet is also invoked in a few blazars with the peculiar observational properties. 1ES 1959+650 is among a few blazars with such peculiar 
characteristics. This source has been a potential object for the case study to explain the broadband SED measured at several occasions using one as well 
as two-zone models in recent times \citep{Acciari2020,Sahu2021,Ghosal2022}. 
\par
1ES 1959+650 is a well studied nearby blazar located at a redshift of $z=0.048$ \citep{Vaucouleurs1991,Perlman1996}. It was first detected in the 
radio at 4.85 GHz and in the X-ray band in early 1990s \citep{Gregory1991,Elvis1992}. With its first detection of the VHE $\gamma$-rays in 1998, 
it belongs to the early group of the TeV emitting blazars \citep{Nishiyama1999,Holder2003}. This source has undergone numerous flaring activities 
at TeV \citep{Aharonian2003,Krawczynski2004,Aliu2014} and X-ray energies with a significant variation in the X-ray spectrum during the enhanced 
emission state \citep{Albert2006,Tagliaferri2008,Aliu2013, Kapanadze2016,Kaur2017,Patel2018,Chandra2021}.The correlation between the X-ray integral flux 
and spectral index suggested a harder-when-brighter behaviour which is an observed feature of most of the blazars \citep{Hayashida2015,Singh2019a}. 
The features derived from long-term multi-wavelength observations \citep{Acero2015,Ajello2017,Patel2018} place 1ES 1959+650 among the extremely 
high synchrotron-peaked blazar candidates with hard $\gamma$-ray spectra \citep{Costamante2018,Singh2019c}. Apart from the unusual emission properties of 
the electromagnetic radiation, 1ES 1959+650 is also considered as a potential blazar candidate source of the astrophysical neutrinos \citep{Halzen2005,Aartsen2019}. 
In the present work, we focus on its long-term optical polarization behaviour using the archival data. The structure of paper is as follows. 
In Section \ref{sec:data}, description of the data set used in this work is given. Results are presented and discussed in detail in Section \ref{sec:result} followed 
by conclusion in Section \ref{sec:summary}. 

%---------------------------Figure:1 Light Curve-----------------------------------------------
\begin{figure}[h]
\centering
\includegraphics[width=0.9\textwidth]{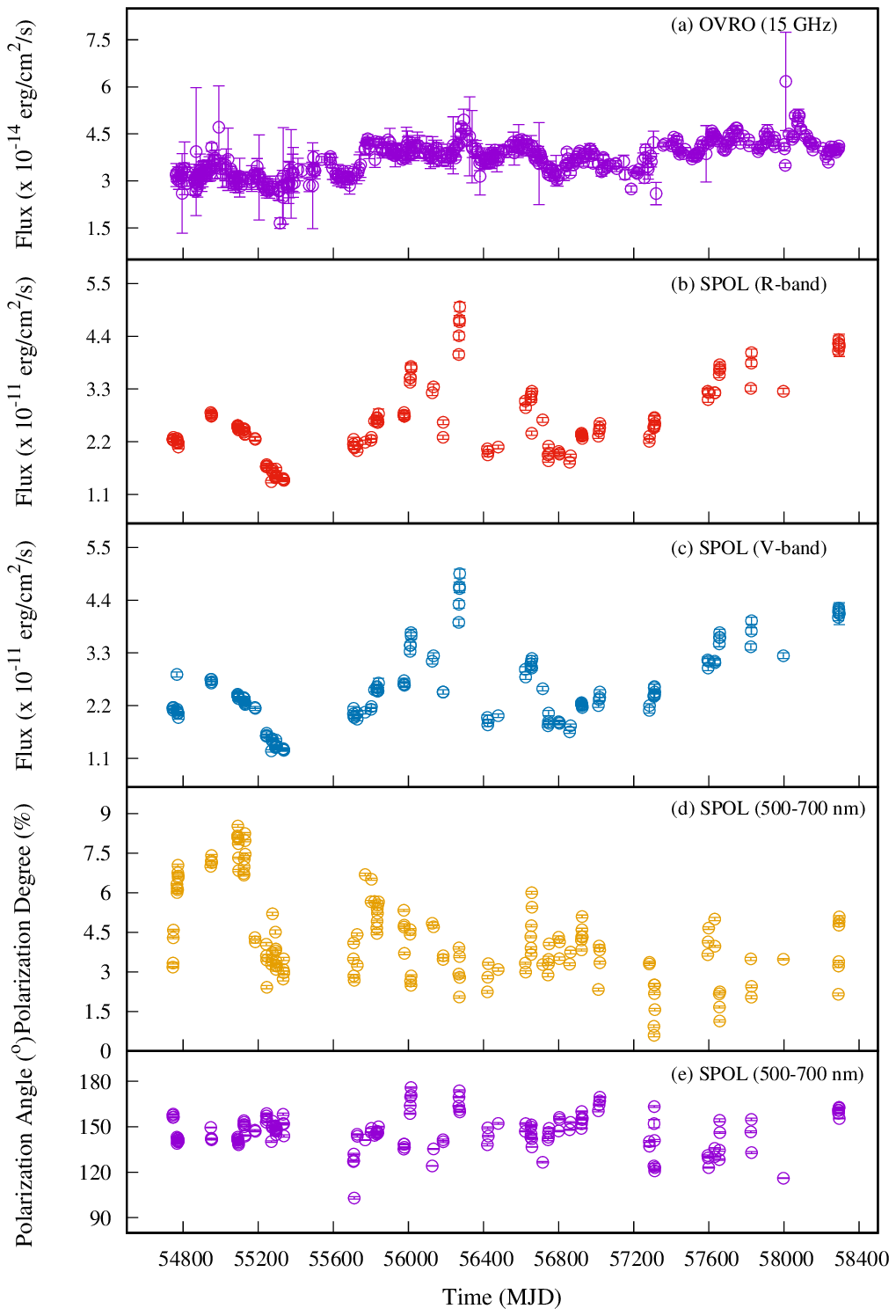}
\caption{Light curve of the blazar 1ES 1959+650 during the period from October 1, 2008 to June 30, 2018 in the radio 
	and optical wavebands measured by the OVRO and the SPOL.} 
\label{lc}
\end{figure}
%---------------------------------------Section-3:Data Set---------------------------------------------------------
\section{Data Set}\label{sec:data}
In this work, we have used the archival data available from the optical and radio observations of the blazar 1ES 1959+650 during the period 
from October 1, 2008 to June 30, 2018 (MJD 54739-58299). These data have been publicly provided under the \emph{Multiwavelength Observing - 
Support Programs}\footnote{https://fermi.gsfc.nasa.gov/ssc/observations/multi/programs.html} to help the \emph{Fermi} science objectives.
Data on the measurements of the optical linear polarization of the source in the wavelength range of $\lambda =$500-700 nm are obtained from the 
Spectro-Polarimeter (SPOL) at the Steward Observatory of the University of Arizona\footnote{http://james.as.arizona.edu/$\sim$psmith/Fermi/DATA/individual.html}.
The SPOL monitoring program has intent to provide photometry and polarimetry measurements of the optical emission from the high energy $\gamma$-ray 
bright blazars \citep{Smith2009}. The polarimeter in the SPOL has a dual beam design incorporating a full aperture Wollaston prism and rotating 
achromatic waveplates \citep{Schmidt1992}. A $\lambda/2$ waveplate is employed to measure the broadband linear polarization at a wavelength resolution 
of $\sim$ 2 nm. Primary data products of the spectropolarimetry are the normalized Stokes parameters q and u with their errors following a normal 
distribution. The degree of optical linear polarization (P) and the corresponding angle of polarization ($\phi$) are given by 
\begin{equation}
\rm P(\%)~=~100 \times \sqrt{q^2 + u^2}
\end{equation}	
and 
\begin{equation}
\rm \phi~=~\frac{1}{2}~~\arctan\left(\frac{u}{q}\right)
\end{equation}	
where the P-values are corrected for the inherent statistical bias in the measurements \citep{Wardle1974}. The measured polarization is rotated to the 
standard astronomical reference frame by monitoring the known interstellar polarization standard stars. A minimum of the two standard stars are observed 
for consistency check during each observation campaign. The polarization angle value increases going East of North. Error in the polarization 
angle is statistical only. In addition to the spectropolarimteric data, we also use the contemporary photometric data from the SPOL measured in the R and V-bands. 
The calibration is performed using the differential flux measurements on the stars in the field of view the source blazar. The measured optical fluxes have been 
corrected for the airmass of the target blazar using the standard extinction curve available within the IRAF reduction package \citep{Smith2009,Stone1983}. 
The optical data in the present study are complemented by the near simultaneous radio measurements at 15 GHz from the Owens Valley 
Radio Observatory (OVRO)\footnote{http://www.astro.caltech.edu/ovroblazars.}. More details of the fast-cadence radio monitoring program of the blazars at 15 GHz 
using the 40m telescope can be found in \citep{Richards2011}.   

%---------------------------------------Section-4:Results-----------------------------------------------------------
\section{Results and Discussion}\label{sec:result}
\subsection{Light curves}\label{sec:lc}
The decade long emission history of the blazar 1ES 1959+650, in the radio and the optical bands for the period October 1, 2008 and June 30, 2018 
(MJD 54739-58299), as a function of time is shown in Figure \ref{lc}(a-e). The data points in the radio light curve (Figure \ref{lc}a) 
indicate an average flux level of (3.86$\pm$0.02)$\times 10^{-14}~erg~cm^{-2}~s^{-1}$ from the source at 15 GHz frequency. 
A visual inspection of the optical light curves (Figure \ref{lc}b,c) suggests that the source emission in the R and V-bands exhibits 
much shorter duration low and high activity states during the whole period considered in this study. A constant fit to the daily flux points 
results in an average flux levels of (2.19$\pm$0.05)$\times 10^{-11}~erg~cm^{-2}~s^{-1}$ and (2.01$\pm$0.05)$\times 10^{-11}~erg~cm^{-2}~s^{-1}$ in the R and V-bands respectively. 
The degree of linear polarization (P) in the wavelength range of 500-700 nm seems to vary in an erratic manner from the visual inspection of the 
light curve depicted in Figure \ref{lc}d. In order to quantify the randomness present in this data, we estimate the well known Shannon entropy \citep{Shannon1948}. 
The derived value of 2.86 for the Shannon entropy indicates a moderate level of randomness present in the temporal behaviour of the degree of linear 
polarization of the  blazar 1ES 1959+650. A constant fit to the polarization measurements yields an average degree of linear polarization 
as $\langle P \rangle~=$ 4.60$\pm$0.15$\%$ with a maximum and minimum measured values of 8.53$\pm$0.05$\%$ and 0.60$\pm$0.06$\%$ respectively. 
This implies that the ratios of the minimum and maximum degrees of linear polarization to the average value are 0.13 and 1.85 respectively. 
Thus, the optical linear polarization of the blazar 1ES 1959+650 has varied significantly in a moderately random way during the period of one 
decade. The polarization angle, reported in Figure \ref{lc}e, is also observed to change with a mean value of $\langle \phi \rangle~=~144^\circ \pm 0.71^\circ$ and 
standard deviation of 11.68. This indicates that the polarization vector is mostly oriented in the East-South (West-North) direction in the sky plane. 
%-----------------------------------Table:1Fvar-----------------------------------------------------
\begin{table}[h]
\caption{Summary of the fractional variability amplitude (F$_{var}$) estimated for the blazar 1ES 1959+650 during the period October 1, 2008 - June 30, 2018 
	(MJD 54739-58299).}\label{Fvar}
\vspace{1.0cm}
\begin{tabular}{lcll}
\toprule
Observable 		&Waveband  	&$F_{var} \pm \Delta F_{var}$\\
\midrule
Radio Flux      	&15GHz        &13.12$\pm$0.60 \%\\
Optical Flux  		&R-band       &30.17$\pm$1.78 \%\\
Optical Flux    	&V-band       &31.87$\pm$1.88 \%\\
Degree of Polarization 	&500-700nm    &38.97$\pm$2.29 \%\\
Angle of Polarization 	&500-700nm    &7.93$\pm$0.46  \%\\
\botrule
\end{tabular}
\end{table}

%---------------------------Figure:2 Correlation-----------------------------------------------
\begin{figure}[h]
\centering
\includegraphics[width=0.6\textwidth,angle=-90]{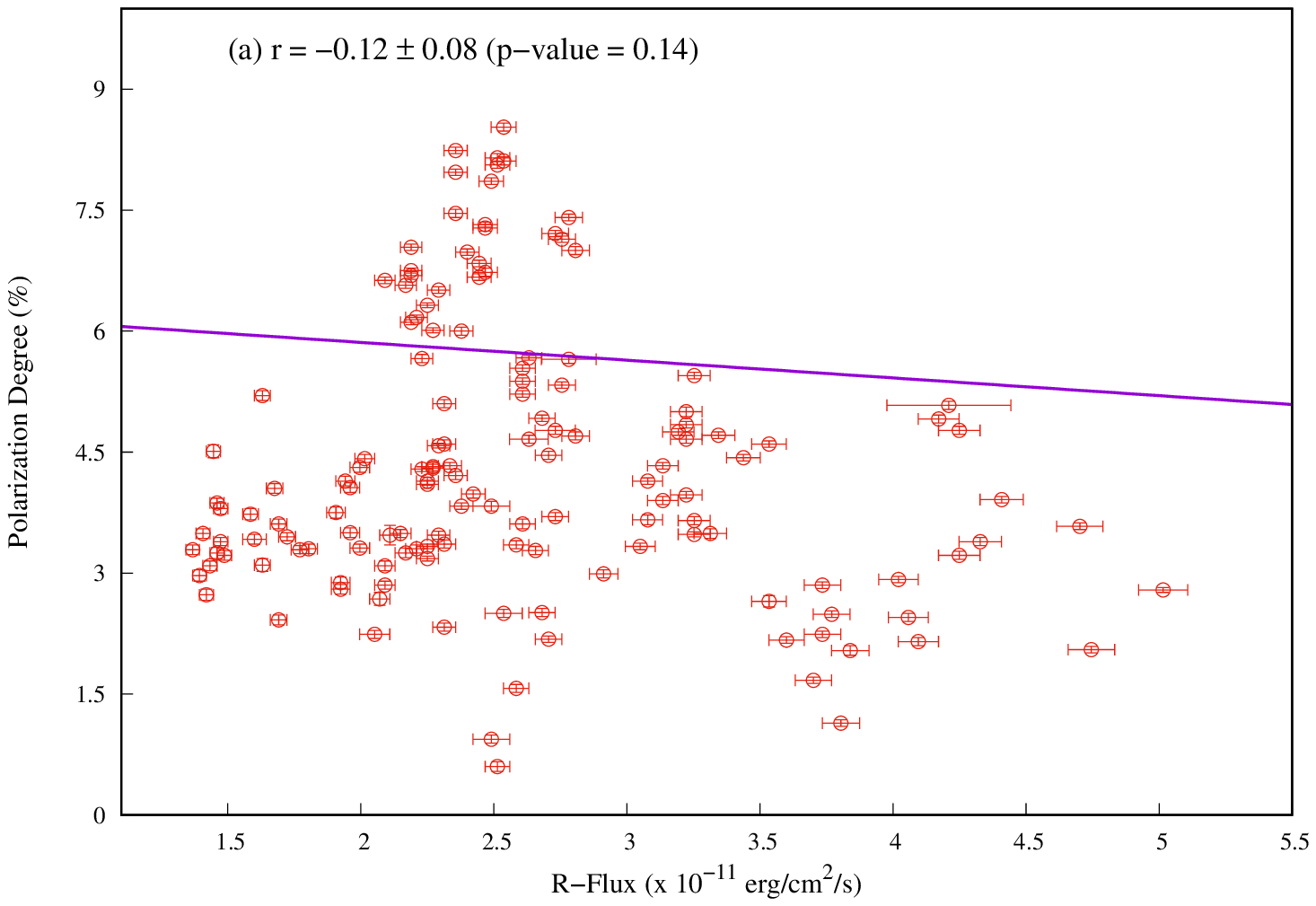}
\includegraphics[width=0.6\textwidth,angle=-90]{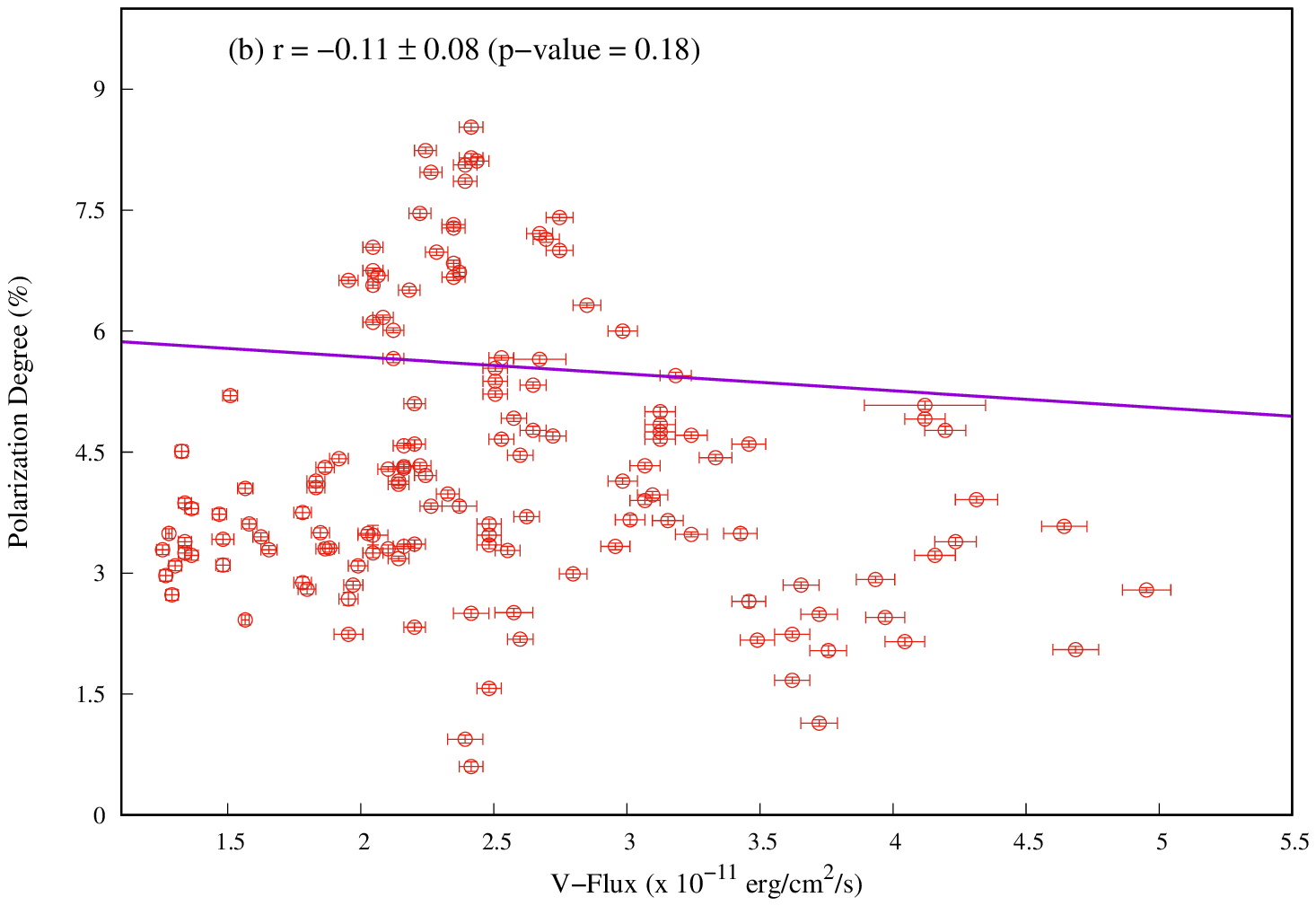}
\caption{Scatter plots for the degree of linear polarization vs R and V-band flux measurements of the blazar 1ES 1959+650.} 
\label{corr}
\end{figure}
%--------------------------------------------------------------------------------------------------------------
\subsection{Variability and Correlations}
As discussed in Section \ref{sec:lc}, the optical emissions measured from the blazar 1ES 1959+650 in the R and V-bands are highly variable 
with the ratio of maximum to minimum flux being $\sim$ 5, whereas the ratio of maximum to minimum flux is $\sim$ 2 for the radio 
emission at 15 GHz. In order to quantify the exact level of intrinsic variability present in the various observables of the source, 
we estimate the so called fractional variability (F$_{\rm var}$) parameter. It is defined as \citep{Vaughan2003,Singh2020b}   
\begin{equation}
\rm F_{var}=\sqrt{\frac{S^2 -E^2}{F^2}}
\end{equation}
and the error in F$_{\rm var}$ is given by 
\begin{equation}
\rm \Delta F_{var}=\sqrt{\left(\sqrt{\frac{1}{2N}}\frac{E^2}{F^2F_{var}}\right)^2+\left(\sqrt{\frac{E^2}{N}}\frac{1}{F}\right)^2}
\end{equation}  
where S$^2$, E$^2$, and F are the variance, the mean square error, and the mean value of the data set, respectively. N is the number of 
data points in the sample. The value of F$_{\rm var}$ accounts for the measurement uncertainties and therefore gives the intrinsic level of 
variability in a data set. If S$^2$~$<$~E$^2$, real value of F$_{\rm var}$ cannot be estimated, indicating that no variability is present in the 
data in addition to the expected noise. And if S$^2$~$>$~E$^2$, F$_{\rm var}$~$>$~3 $\Delta$F$_{var}$ suggests a significant variability as per 
the convention \citep{Otsuki2024}. The estimated values of  F$_{\rm var}$ $\pm$ $\Delta$F$_{var}$ for all the data sets reported in Figure \ref{lc} are 
shown in Table \ref{Fvar}. It is evident that the intrinsic flux variability in the optical R and V-bands can be quantified by higher  F$_{\rm var}$ values 
than that of the radio emission at 15 GHz. This indicates that the optical emission is more variable than the radio emission in the blazar 
1ES 1959+650 observed over a decade. The degree of linear polarization exhibits highest value of F$_{\rm var}$, indicating a significantly variable 
polarized emission from the source in the optical band. As the flux points in the R and V-bands are found to be significantly variable, we look for a 
possible correlation with the degree of linear polarization. For this, we estimate the Pearson's correlation coefficient ($r$) and corresponding p-values for 
the scatter plots between the near simultaneous measurements of the polarization degree and the R and V-band fluxes reported in Figure \ref{corr}(a-b). 
The estimated values of the Pearson's coefficient ($r~\sim~-0.12$) with a p-value of $\sim$ 0.1 indicate that there is no significant correlation 
for the degree of linear polarization with the optical emissions in the R and V-bands.This can be attributed to the multi-zone emission 
scenario \citep{Marscher2008} in the jet of the blazar 1ES 1959+650. The polarized emission may originate from a local shocked region wherein 
the tangled magnetic field lines get partially ordered due to the shock propagation inside the jet filled with the turbulent plasma. The compression 
of plasma due to the shock propagation results in the acceleration of electrons to the relativistic energies \citep{Marscher2014}. The maximum 
energy of the accelerated electrons depends on the orientation of the magnetic field with respect to the shock front. This leads to an overall 
increase in the optical emission by the synchrotron radiation of electrons, whereas the net degree of linear polarization may decrease due to 
the partial cancellations produced by the multiple emission zones having the similar magnetic field strengths and random orientations. Therefore, 
the optical flux and the degree of linear polarization vary due to continuous perturbations caused by the presence of turbulence in the relativistic 
jets. However, the angle of polarization may not change gradually due to the compression of plasma by a perpendicular shock moving along the matter 
dominated jet. Alternatively, the blazar jets with an ordered helical magnetic field and varying speed also cause random changes in the degree 
of polarization and associated physical properties \citep{Lyutikov2017}. The regularly changing jet orientation and possibly a changing Doppler boosting 
can produce the variable degree of polarization and swings in the angle of polarization through a highly deterministic process.
\par
Theoretically, the intrinsic degree of linear polarization produced by the synchrotron emission of the relativistic electrons with a power law energy spectrum 
is given by \citep{Roux1961,Rybicki1986}: 
\begin{equation}\label{pol}
\rm	P~=\frac{s~+~1}{s~+~7/3}.
\end{equation}	
where s is the power law spectral index of the electron energy distribution. Assuming a typical value of s~=~2-3 for the blazars, the maximum intrinsic degree 
of linear polarization (P$_{\rm max}$) is expected to be P$_{\rm max}$~=~ 69-75\%. If the pattern of the turbulence is approximated by the cells of uniform 
magnetic field with random orientations in the jet, the mean value of the degree of linear polarization is expressed as \citep{Marscher2014}
\begin{equation}
\rm  \langle P \rangle~=~\frac{P_{max}}{\sqrt{N_0}}
\end{equation}
where N$_{\rm 0}$ is the number of magnetic cells representing the turbulence in the jet. For $\langle P \rangle~=0.046$ as discussed in Section \ref{sec:lc} and 
assuming P$_{\rm max}$~=~0.7, the number of magnetic cells is estimated as N$_{\rm 0}$~=~231. This means that 230 cells with uniform magnetic fields of random 
orientations are required to produce the mean value ($\langle P \rangle~=0.046$) of the observed degree of linear polarization from the blazar 1ES 1959+650. 
It is worth mentioning that the mean value $\langle P \rangle~=0.046$ does not represent the true mean value of the intrinsic degree of polarization due to 
several depolarization effects which will be discussed later in detail in Section \ref{model}.  
%---------------------------Figure:3 Modelling-----------------------------------------------
\begin{figure}[h]
\includegraphics[width=0.8\textwidth,angle=-90]{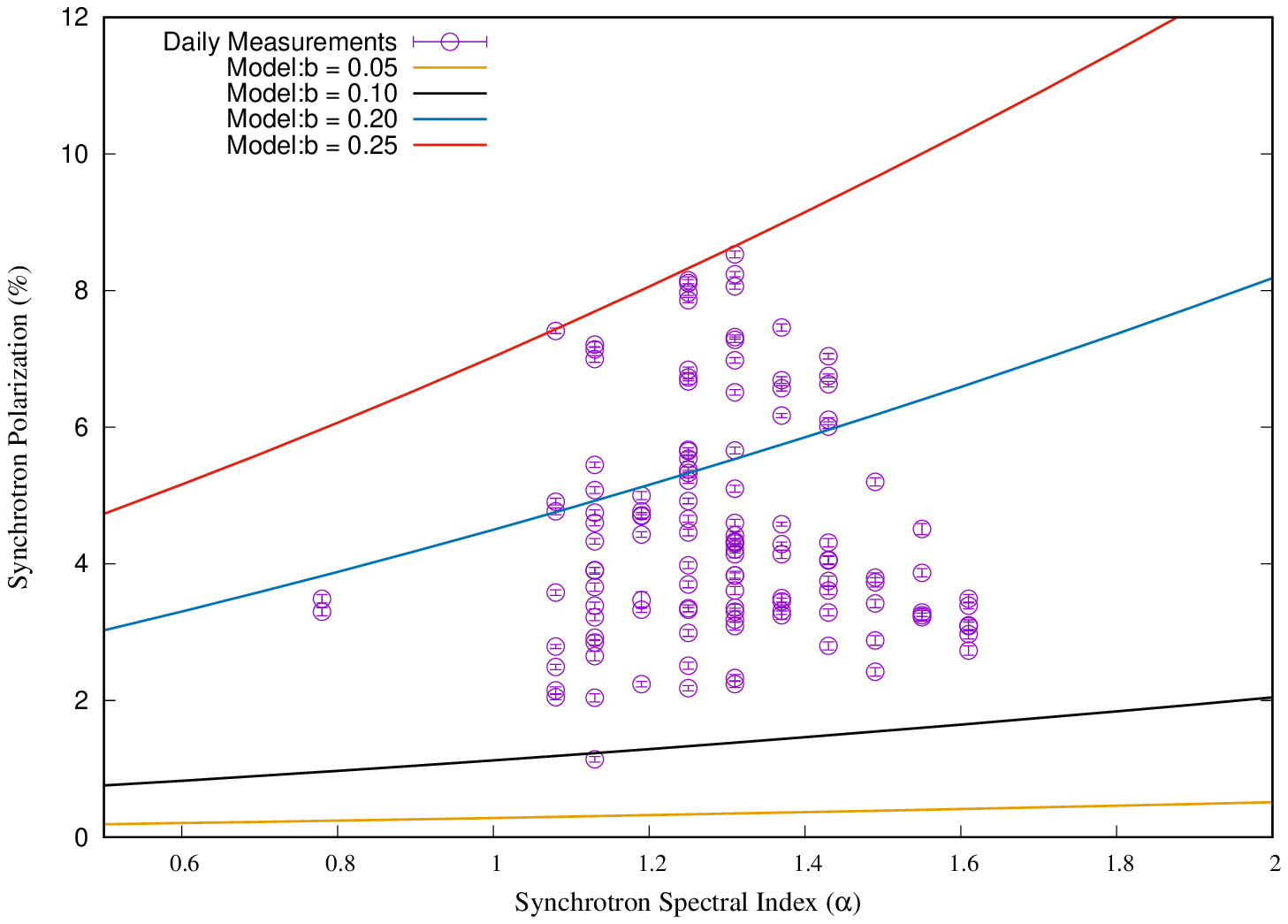}
\caption{The degree of linear polarization as a function of the energy spectral index ($\alpha$) of photons produced by the synchrotron emission. 
	Curves represent the intrinsic degree of polarization for different configurations of the magnetic field. Data points are the measured degree of  
	polarization averaged over one-day during the period from October 1, 2008 to June 30, 2018 (MJD 54739-58299) in the wavelength range of 500-700 nm from 
	the blazar 1ES 1959+650.} 
\label{mod}
\end{figure}
%----------------------------------------------------------------------------
\subsection{Polarization Modelling}\label{model}
It is widely accepted that the synchrotron emission of the relativistic electron-positron population, moving in a large scale and ordered magnetic field 
(e.g. blazar jets), produces a strongly polarized electromagnetic radiation in the radio and optical bands. The electrons are generally accelerated to the 
relativistic energies by the putative physical processes such as the \emph{Fermi} first and second order acceleration via diffusive shocks and stochastic 
turbulences respectively \citep{Zheng2018} and the magnetic reconnection \citep{Sironi2015a}. The \emph{Fermi} acceleration mechanism (both first and second order) 
usually produces an electron energy distribution described by a simple power law \citep{Qin2018}. For an isotropic distribution of the relativistic electrons 
following a power law energy spectrum, the emitted synchrotron radiation is also described by a power law of the form \citep{Rybicki1986}
\begin{equation}\label{syn-spec} 
\rm 	F_\nu~\propto~\nu^{-\alpha}
\end{equation}	
where $\rm \nu$ is the frequency of radiation and $\rm \alpha$=(s~-1)/2 is the synchrotron spectral index. If the optical emission in the R and V-bands is attributed 
to the synchrotron process, then from Equation \ref{syn-spec}, we have 
\begin{equation}\label{spec-index}
\rm 	\alpha~=~ \frac{log\left(\frac{F_\nu ^R}{F_\nu ^V}\right)}{log\left(\frac{\nu_V}{\nu_R} \right)}
\end{equation}
where \rm F$_\nu ^R$ and \rm F$_\nu ^V$ are the flux densities at the central frequencies $\rm \nu_R$ and $\rm \nu_V$ in the R and V-bands respectively. 
For $\rm \nu_R~= 4.68\times10^{14}$ Hz and $\rm \nu_V~= 5.47\times10^{14}$ Hz, Equation \ref{spec-index} can be written as 
\begin{equation}
\rm 	\alpha~=~14.76~log\left(\frac{F_\nu ^R}{F_\nu ^V}\right)
\end{equation}
Thus, from the measured values of $\rm \nu_R F_\nu ^R$ and $\rm \nu_V F_\nu ^V$ ($\rm \nu F_\nu$ representation in the SED), the synchrotron spectral index 
can be estimated using optical observations in the R and V-bands (Figure \ref{lc}). We pair the derived values of $\alpha$ with the corresponding 
observed P-values (as shown in Figure \ref{mod}) and find that the synchrotron spectral index ranges between $\rm 0.8 < \alpha < 1.6$. This implies 
that the energy distribution of particles or electrons, producing synchrotron radiation in the jet of blazar 1ES 1959+650, can be described by a 
power law with spectral index $\rm s~=~2.6-4.2$,which is broadly consistent with the turbulent diffusive shock acceleration scenario \citep{Sironi2015b}.  
Thus, the turbulent shock along with the other putative acceleration mechanisms can be responsible for the acceleration of electrons 
to the relativistic energies in the jet of the blazar 1ES 1959+650. 
\par
In the frame-work of a two zone emission model, the superposition of a constant polarization originating from one zone and a variable polarization from 
second zone is invoked to understand the observed polarization from blazars during their different activity states \citep{Fraija2017}. Variation in the 
degree of polarization is attributed to the evolution of magnetic field lines configuration in the emission region \citep{Korchakov1962,Bjornsson1985}.
For an optically thin synchrotron emission region, the degree of linear polarization  depends primarily on the uniform magnetic field structure 
(configuration of the prevalent magnetic field lines) and partially on the electron energy distribution emitting the synchrotron radiation. In a simple scenario, 
we assume that the resultant magnetic field (B) in the emission region is produced by an ordered ($\rm B_o$) and a chaotic ($\rm B_c$) magnetic field. For such a
magnetic field structure, the intrinsic degree of linear polarization is given by \citep{Fraija2017,Singh2019b}
\begin{equation}\label{pol-mod}
 \rm 	P_{int}~=~\frac{(\alpha + 1)(\alpha + 2)(\alpha + 3)}{8(\alpha + 5/3)}b^2
\end{equation}
where $\rm b=B_o/B_c$. The variation of $\rm P_{int}$ as a function of $\rm \alpha$ for the different configurations of $\rm B_o$ and $\rm B_c$ in 
the synchrotron emission region is presented in Figure \ref{mod}. It is evident that the degree of polarization increases drastically at a given 
$\rm \alpha$ if the ordered magnetic field starts dominating over the chaotic magnetic field. In the asymptotic case of a perfectly ordered uniform 
magnetic field configuration, Equation \ref{pol-mod} approaches to Equation \ref{pol} and gives the maximum degree of polarization of $\sim 70\%$ for 
$\rm 0.5~<~\alpha~<~1$. On the other hand, a completely isotropic chaotic magnetic field produces negligible polarization. The minimum and maximum degrees of 
polarization measured from the blazar 1ES 1959+650 correspond to an ordered magnetic field of $\sim~10\%$ and $\sim~25\%$ of the chaotic magnetic field respectively 
in the emission region. This plausibly suggests that compression of the chaotic magnetic field by a moving shock produces a partially ordered magnetic field which 
leads to the polarized emission from the jet. Thus, the random behaviour of the observed degree of polarization from the source over a decade can be attributed 
to the change in electron energy distribution and relative ordering of the magnetic field caused by a shock compression in the emission region. 
It has also been shown that change in the bulk speed of the emission region (or bulk Lorentz factor) and the viewing angle can lead to random 
changes in the properties of the synchrotron polarization under a highly ordered helical magnetic field in the emission region \citep{Lyutikov2017}. 
\par
The values of the observed degree of polarization (data points in Figure \ref{mod}) can be lower than the corresponding intrinsic values due to the several 
depolarization effects including contamination of the stellar thermal emission from the host galaxy and inhomogeneties in the magnetic field \citep{Netzer2013}. 
Effect of the non-polarized stellar emission becomes significant when the host galaxy is brighter than the jet emission and a large photometric aperture is used for 
polarization measurements \citep{Andruchow2008}. The host galaxy of the blazar 1ES 1959+650 is well resolved 
and has a large angular diameter \citep{Urry2000}. Therefore, the depolarization effect of the host galaxy on the intrinsic polarization can be investigated 
for this source. In order to correct for the depolarization effects introduced by the host galaxy of the source, we use the following relation between the intrinsic 
or jet polarization ($\rm P_{jet}$) and the measured polarization ($\rm P_{obs}$) given in \citep{Carnerero2017}:
\begin{equation}
 \rm	P_{jet}~=~\frac{P_{obs}}{1-\frac{F_{host}}{F_{obs}}}
\end{equation}	
where $\rm F_{host}$ is the flux of the host galaxy and $\rm F_{obs} (~=~F_{jet} + F_{host})$ is the total observed flux involving the contributions from 
the blazar jet ($\rm F_{jet}$) as well as the host galaxy. For the blazar 1ES 1959+650, the host is a bright elliptical galaxy with 
$\rm F_{host}~\sim~8.09\times 10^{-12}~erg~cm^{-2}~s^{-1}$ in the R-band \citep{Scarpa2000}. Assuming the value of $\rm F_{jet}$ to be an average value 
of the R-band flux as $\rm F_{jet}~\sim~2.20\times10^{-11}~erg~cm^{-2}~s^{-1}$ from Section \ref{sec:lc}, we find that $\rm P_{jet}~=1.36~\times~P_{obs}$. 
This indicates that the minimum and maximum values of the intrinsic or jet polarization for the blazar 1ES 1959+650 will at least be $\sim$ 0.82\% and 12\% in 
the wavelength range of 500-700 nm. These values may further increase after accounting for the depolarization due to inhomogeneous jet magnetic field, which 
is beyond the scope of this work.  Therefore, 1ES 1959+650 can be classified as a high-polarization blazar of BL Lacertae 
type according to the opto-polarimetric division of the quasi-stellar objects \citep{Stockman1984}. In a recent study, a higher degree of linear polarization 
(up to $\sim$ 8\%) at the X-ray energies in the range 2-8 keV has been measured from this source by the Imaging X-ray Polarimetry Explorer \citep{Errando2024}. 
Also, an upper limit of 5.1\% on the X-ray polarization is found to be comparable to the optical polarization at a different epoch. Authors have attributed 
the X-ray polarization of the source to the turbulence in the jet plasma flow. The highest degree of the X-ray polarization measured from 1ES 1959+650 in the energy 
range of 2-8 keV is $\sim$ 13\% \citep{Hu2024}. The X-ray polarization measurements from the source at different epochs are attributed to the shock acceleration 
of electrons to the relativistic energies in the jet. The log-normal distribution of the X-ray fluxes in the long-term light curve of 1ES 1959+650 also supports 
the hypothesis of the relativistic shocks propagating down the jet \citep{Wani2023}. The formation of these shocks can be related to the inhomogeneties or turbulences 
occurring in the accretion disk. Inward propagating fluctuations in the accretion disk from different radii cumulate in a multiplicative way \citep{Arvalo2006}. The 
log-normal distribution, attributed to the underlying multiplicative physical processes originating in the accretion disk, might be an indication of the variability 
imprint of the accretion disk onto the jet \citep{Uttley2005}. The above argument assumes that the fluctuations in the accretion disk are efficiently transmitted to the jet, 
requiring a standard accretion disk and its connection with the jet. The flux variability at the short time scales is triggered by the interaction of the shock 
front with the jet inhomogenities.  
 
%-----------------------------------Section:6-Conclusion----------------------------------------
\section{Conclusion}\label{sec:summary}
The degree of linear polarization in the wavelength range of 500-700 nm from the blazar 1ES 1959+650 ranges between $\sim$ 0.6\% to $\sim$ 8.5\% during 
a period of about ten years considered in this study. Long-term behaviour of the polarization is found to be very complex and more variable than the optical 
emission in the R and V-bands. The angle of polarization tends to be relatively stable with a preferable orientation along the East-South (West-North) 
direction in the sky plane. A weak anti-correlation between degree of linear polarization and optical fluxes in the R and V-bands hints at the multi-component 
emission in the shock-in-jet scenario for the blazar 1ES 1959+650. A turbulent shock moving down the jet may create a maximum of $\sim$ 230 synchrotron emitting 
shells within a single zone to produce the observed variable nature of the polarization. The synchrotron spectral index is observed to vary between 0.8 to 1.6. 
The synchrotron emission region in the jet is embedded with a complex configuration of magnetic field having ordered and chaotic components. Dominance of the 
one component over the other due to the shock compression can lead to variability in the polarized emission. The maximum degree of linear polarization corresponds 
to an ordered magnetic field strength of about 25\% of the chaotic magnetic field in the emission region. A comparison of the average values of the degree 
of linear polarization before and after accounting for the depolarization effects induced by the host galaxy suggests that the jet polarization should be significantly 
higher than the measured values at least by a factor $\sim$ 1.36. Therefore, the contamination of the stellar thermal emission from the host galaxy of 1ES 1959+650 
strongly affects the synchrotron polarization in the jet. Simultaneous measurements of the polarization over a wide waveband from the optical to the X-rays and the 
high $\gamma$-rays using future instruments are essential to probe the exact physical processes involved in the non-thermal emission from the blazar jets.

%------------------------------------------------------------------------------------------------------
\section*{Data Availability}
Data used in this work can be downloaded from http://james.as.arizona.edu/$\sim$psmith/Fermi/DATA/individual.html.

%-------------------------------------------------------------------
\section*{Declaration of competing interest}
The authors declare that they have no known competing financial interests or personal relationships that could have appeared to
influence the work reported in this paper.

%-------------------------------------Acknowledgements----------------------------------------------
\bmhead{Acknowledgements}
We thank the anonymous reviewer for his/her important and constructive suggestions which have greatly helped in improving the contents of the manuscript
scientifically. 
Data from the Steward Observatory spectropolarimetric monitoring project were used. This program is supported by 
Fermi Guest Investigator grants NNX08AW56G, NNX09AU10G, NNX12AO93G, and NNX15AU81G.
This research has made use of data from the OVRO 40-m monitoring program (Richards, J. L. et al. 2011, ApJS, 194, 29) which 
is supported in part by NASA grants NNX08AW31G, NNX11A043G, and NNX14AQ89G and NSF grants AST-0808050 and AST-1109911.

%%---------------------------------------------References-----------------------------------------
\bibliography{MS}% common bib file
%%\input MS.bbl

\end{document}